\documentclass[twoside,twocolumn,9pt]{article}
\usepackage{extsizes}
\usepackage[super,sort&compress,comma]{natbib} 
\usepackage[version=3]{mhchem}
\usepackage[left=1.5cm, right=1.5cm, top=1.785cm, bottom=2.0cm]{geometry}
\usepackage{balance}
\usepackage{mathptmx}
\usepackage{sectsty}
\usepackage{graphicx} 
\usepackage{lastpage}
\usepackage[format=plain,justification=justified,singlelinecheck=false,font={stretch=1.125,small,sf},labelfont=bf,labelsep=space]{caption}
\usepackage{float}
\usepackage{fancyhdr}
\usepackage{fnpos}
\usepackage[english]{babel}
\usepackage{siunitx}
\addto{\captionsenglish}{%
  \renewcommand{\refname}{Notes and references}
}
\usepackage{array}
\usepackage{droidsans}
\usepackage{charter}
\usepackage[T1]{fontenc}
\usepackage[usenames,dvipsnames]{xcolor}
\usepackage{setspace}
\usepackage[compact]{titlesec}
\usepackage{hyperref}
\usepackage{footnote}
\usepackage{tablefootnote}
\makesavenoteenv{tabular}
\definecolor{cream}{RGB}{222,217,201}

\begin{document}

\pagestyle{fancy}
\thispagestyle{plain}
\fancypagestyle{plain}{
%%%HEADER%%%
\renewcommand{\headrulewidth}{0pt}
}
%%%END OF HEADER%%%

%%%PAGE SETUP - Please do not change any commands within this section%%%
\makeFNbottom
\makeatletter
\renewcommand\LARGE{\@setfontsize\LARGE{15pt}{17}}
\renewcommand\Large{\@setfontsize\Large{12pt}{14}}
\renewcommand\large{\@setfontsize\large{10pt}{12}}
\renewcommand\footnotesize{\@setfontsize\footnotesize{7pt}{10}}
\renewcommand\scriptsize{\@setfontsize\scriptsize{7pt}{7}}
\makeatother

\renewcommand{\thefootnote}{\fnsymbol{footnote}}
\renewcommand\footnoterule{\vspace*{1pt}% 
\color{cream}\hrule width 3.5in height 0.4pt \color{black} \vspace*{5pt}} 
\setcounter{secnumdepth}{5}

\makeatletter 
\renewcommand\@biblabel[1]{#1}            
\renewcommand\@makefntext[1]% 
{\noindent\makebox[0pt][r]{\@thefnmark\,}#1}
\makeatother 
\renewcommand{\figurename}{\small{Fig.}~}
\sectionfont{\sffamily\Large}
\subsectionfont{\normalsize}
\subsubsectionfont{\bf}
\setstretch{1.125} %In particular, please do not alter this line.
\setlength{\skip\footins}{0.8cm}
\setlength{\footnotesep}{0.25cm}
\setlength{\jot}{10pt}
\titlespacing*{\section}{0pt}{4pt}{4pt}
\titlespacing*{\subsection}{0pt}{15pt}{1pt}
%%%END OF PAGE SETUP%%%

%%%FOOTER%%%
\fancyfoot{}
\fancyfoot[LE]{\footnotesize{\sffamily{\thepage~\textbar\hspace{3.45cm} 1--\pageref{LastPage}}}}
\fancyhead{}
\renewcommand{\headrulewidth}{0pt} 
\renewcommand{\footrulewidth}{0pt}
\setlength{\arrayrulewidth}{1pt}
\setlength{\columnsep}{6.5mm}
\setlength\bibsep{1pt}
%%%END OF FOOTER%%%

%%%FIGURE SETUP - please do not change any commands within this section%%%
\makeatletter 
\newlength{\figrulesep} 
\setlength{\figrulesep}{0.5\textfloatsep} 

\newcommand{\topfigrule}{\vspace*{-1pt}% 
\noindent{\color{cream}\rule[-\figrulesep]{\columnwidth}{1.5pt}} }

\newcommand{\botfigrule}{\vspace*{-2pt}% 
\noindent{\color{cream}\rule[\figrulesep]{\columnwidth}{1.5pt}} }

\newcommand{\dblfigrule}{\vspace*{-1pt}% 
\noindent{\color{cream}\rule[-\figrulesep]{\textwidth}{1.5pt}} }

\makeatother
%%%END OF FIGURE SETUP%%%

%%%TITLE AND AUTHORS%%%
\twocolumn[
  \begin{@twocolumnfalse}
\vspace{1em}
\sffamily

 \noindent\LARGE{\textbf{Laboratory blueprints for interstellar searches of aromatic chiral molecules: rotational signatures of styrene oxide$^\dag$}} \\
  \vspace{0.3cm} \\

  \noindent\large{Pascal Stahl,$^{\ast}$\textit{$^{a}$} Benjamin E. Arenas,\textit{$^{b}$} Sérgio R. Domingos,\textit{$^{b}$} Guido W. Fuchs,\textit{$^{a}$} Melanie Schnell,\textit{$^{bc}$} and Thomas F. Giesen\textit{$^{a}$}} \\%Author names go here instead of "Full name", etc.

 \end{@twocolumnfalse} \vspace{0.6cm}

  ]
%%%END OF TITLE AND AUTHORS%%%

%%%FONT SETUP - please do not change any commands within this section
\renewcommand*\rmdefault{bch}\normalfont\upshape
\rmfamily
\section*{}
\vspace{-1cm}

%%%FOOTNOTES%%%

\footnotetext{\textit{$^{a}$Institute of Physics, University of Kassel, Heinrich-Plett-Str. 40, 34132 Kassel, Germany; E-mail: p.stahl@physik.uni-kassel.de}}
\footnotetext{\textit{$^{b}$Deutsches Elektronen-Synchrotron (DESY), Notkestr. 85, 22607 Hamburg, Germany. }}
\footnotetext{\textit{$^{c}$Institut für Physikalische Chemie, Christian-Albrechts-Universität zu Kiel, Max-Eyth-Str. 1,
24118 Kiel, Germany. }}

%Please use \dag to cite the ESI in the main text of the article.
%If you article does not have ESI please remove the the \dag symbol from the title and the footnotetext below.
\footnotetext{\dag~Electronic Supplementary Information (ESI) available: Tables S1-S4 and a line list as a .cat file. See DOI: 00.0000/00000000.}
%additional addresses can be cited as above using the lower-case letters, c, d, e... If all authors are from the same address, no letter is required
%[details of any supplementary information available should be included here]
%\footnotetext{\ddag~Additional footnotes to the title and authors can be included \textit{e.g.}\ `Present address:' or `These authors contributed equally to this work' as above using the symbols: \ddag, \textsection, and \P. Please place the appropriate symbol next to the author's name and include a \texttt{\textbackslash footnotetext} entry in the the correct place in the list.}

%%%END OF FOOTNOTES%%%

%%%ABSTRACT%%%%

\sffamily{\textbf{The tracking of symmetry-breaking events in space is a long-lasting goal of astrochemists, aiming at an understanding of homochiral Earth chemistry. One current effort at this frontier aims at the detection of small chiral molecules in the interstellar medium. For that, high-resolution laboratory spectroscopy data is required, providing blueprints for the search and assignment of these molecules using radioastronomy. Here, we used chirped-pulse Fourier transform microwave and millimeter-wave spectroscopy and frequency modulation absorption spectroscopy to record and assign the rotational spectrum of the chiral aromatic molecule styrene oxide, $\mathrm{C_{6}H_{5}C_{2}H_{3}O}$, a relevant candidate for future radioastronomy searches. Using experimental data from the 2-12, 75-110, 170-220, and 260-330 GHz regions, we performed a global spectral analysis, complemented by quantum chemistry calculations. A global fit of the ground state rotational spectrum was obtained, including rotational transitions from all four frequency regions. Primary rotational constants as well as quartic and sextic centrifugal distortion constants were determined. We also investigated vibrationally excited states of styrene oxide, and for the three lowest vibrational states, we determined rotational constants including centrifugal distortion corrections up to the sextic order. In addition, spectroscopic parameters for the singly-substituted $\mathrm{^{13}C}$ and $\mathrm{^{18}O}$ isotopologues were retrieved from the spectrum in natural abundance and used to determine the effective ground state structure of styrene oxide in the gas phase. The spectroscopic parameters and line lists of rotational transitions obtained here will assist future astrochemical studies of this class of chiral organic molecules.
}\\
%\sffamily{\textbf{The abstract should be a single paragraph which summarises the content of the article. Any references in the abstract should be written out in full \textit{e.g.} [Surname \textit{et al., Journal Title}, 2000, \textbf{35}, 3523].}}\\%The abstrast goes here instead of the text "The abstract should be..."

%%%END OF ABSTRACT%%%%

\rmfamily %Please do not remove this line.

%%%MAIN TEXT%%%%

%\input{introduction}
\section{Introduction}
The recent astronomical detection of methyl oxirane ($\mathrm{CH_{3}C_{2}H_{3}O}$) in the star forming region Sagittarius B2(N) has raised renewed interest in simple chiral molecules and triggered laboratory spectroscopic investigations. The observation of chiral molecules in space is intimately linked to the question of the origin of homochirality, which is the preference for left-handed amino acids and right-handed sugars in living organisms. It is a long-standing debate about whether this break of symmetry is purely accidental or has a deeper physical meaning \cite{Blackmond.2010, Saito.2013}. In the course of biological evolution, the enhancing effect of catalytic self-reproduction in living systems may have significantly increased an initially very small difference between left- and right-handed molecules. Such small differences could possibly already result from an interstellar chemical evolution that preceded biological evolution \cite{Huck.1996}. It is therefore crucial to understand the formation of chiral molecules in space and to further investigate their properties through dedicated laboratory experiments.
\\
In general, complex organic molecules (COMs) consisting of several carbon atoms are predominantly found in cold, dense regions of the interstellar medium (ISM) and in the warm gas surrounding young stellar objects \citep{Herbst.2009b}. Current chemical models explain the formation of COMs from small molecules (such as $\mathrm{CO}$, $\mathrm{CO_2}$, $\mathrm{H_{2}O}$, $\mathrm{NH_{3}}$, $\mathrm{CH_4}$) trapped in the icy surfaces of interstellar dust grains. The  energy release of UV-radiation penetrating the ice mantles, the bombardment by electrons and atoms, and processes induced by cosmic rays can initiate chemical reactions, which lead to the formation of COMs from small molecular precursors. Volatile products of these reactions are released from the icy grain surface into space, where they can be observed and assigned by means of radioastronomy. These spectroscopic data provide valuable hints to the hidden processes and the physical conditions that govern the formation of complex molecules \citep{Herbst.2009b}. Chemical models including large sets of gas-grain reactions have been developed to explain the presence of COMs in a qualitative and quantitative manner \citep{Belloche.2009,Garrod.2013,Herbst.2017}. 
\\
The recent discoveries of ethyl formate ($\mathrm{HCOOC_{2}H_{5}}$), \textit{n}-propyl cyanide (\textit{n}-$\mathrm{C_{3}H_{7}CN}$), and the aromatic ring molecule benzonitrile (\textit{c}-$\mathrm{C_{6}H_{5}CN}$) speak for the high degree of astrochemical complexity \citep{Belloche.2009,McGuire.2018b}. 
Particularly, great attention was paid to the detection of \textit{i}-propyl cyanide (\textit{i}-$\mathrm{C_{3}H_{7}CN}$), the first branched alkyl chain \citep{Belloche.2014}, and oxirane (\textit{c}-$\mathrm{C_{2}H_{4}O}$), one of the few ring molecules found in the interstellar medium \citep{Dickens.1997}.
In 2016, McGuire et al. used the 100m Robert C. Byrd Green Bank Telescope (GBT) to record spectral lines of methyl oxirane ($\mathrm{CH_{3}C_{2}H_{3}O}$), the first chiral molecule detected in space \citep{McGuire.2016b}. The rotational spectrum of methyl oxirane was studied up to 40 GHz by Swalen and Herschbach \cite{Swalen.1957, Herschbach.1958}. The most up-to-date laboratory study on its ground state including assignments up to 1 THz was performed by Mesko \textit{et al.} \citep{Mesko.2017}, which also describes the internal rotational motion of this molecule.
\\
Triggered by the recent detections of methyl oxirane and benzonitrile, we investigated the rotational spectra of styrene oxide (SO, $\mathrm{C_{6}H_{5}C_{2}H_{3}O}$), an oxirane with a phenyl ring attached to one of the carbon atoms. SO is a stable chiral molecule with a significant permanent dipole moment of 1.8 Debye. With three strong dipole moment components, $\mu_{a}$=0.6 D, $\mu_{b}$=1.0 D and $\mu_{c}$=1.3 D, and a vapour pressure of 40 Pa at room temperature, it is also perfectly suited for microwave three-wave mixing (M3WM) studies \cite{Patterson.2013,Patterson.2013b,Shubert.2014,V.AlvinShubert.2014,Lobsiger.2015,Domingos2018}, which allow spectroscopic differentiation of R- and S-enantiomers.
In the course of our present investigations, we analysed its vibrational ground state spectrum as well as the rotational transitions originating from its three normal modes $v_{45}$, $v_{44}$, and $v_{43}$. In addition to the main isotopologue, we also studied the spectra of $\mathrm{^{13}C}$- and $\mathrm{^{18}O}$-singly-substituted isotopologues in natural abundance, and we used them to determine precise bond lengths and angles of the ground state gas-phase structure of SO. The high-resolution spectra of SO and the determined rotational constants will enable a dedicated search for this chiral molecule in the ISM.

\section{Experiments and quantum chemical calculations}
\subsection{Experiments}
The rotational spectrum of styrene oxide was recorded over four frequency regions. Figure \ref{fig:bands} shows an overview of the different regions studied and how they overlap with the Effelsberg telescope \footnote[3]{Max Planck Gesellschaft 100m Effelsberg Teleskop, \url{https://eff100mwiki.mpifr-bonn.mpg.de/doku.php?id=information_for_astronomers:rx_list}, (accessed February 2020)}, GBT \footnote[4]{100m Robert C. Byrd Green Bank Telescope, \url{https://greenbankobservatory.org/science/telescopes/gbt/}, (accessed February 2020) }, and Atacama Large Millimeter/submillimeter Array (ALMA) operating bands \footnote[5]{ESO ALMA Receiver Bands, \url{https://www.eso.org/public/teles-instr/alma/receiver-bands/}, (accessed February 2020)}. The spectrometers employed in this study are located at the Deutsches Elektronen-Synchrotron (DESY) facility, where we used the Hamburg COMPACT Spectrometer and a segmented millimeter-wave spectrometer from BrightSpec, Inc., as well as at the University of Kassel, where we used the Kassel THz spectrometer. The instruments in Hamburg are chirped-pulse Fourier transform microwave (CP-FTMW) and millimeter-wave (CP-FTmmw) spectrometers and cover the regions 2-18 GHz and 75-110 GHz, respectively. The Kassel THz spectrometer is a frequency modulation absorption spectrometer operating between 75 GHz and about 1 THz. The measurements conducted for this work cover the 170-220 GHz and 260-330 GHz regions. The instruments have been described in detail previously (references below), and they are introduced briefly in the following sections. In Table \ref{tab:overview}, an overview of the covered spectral ranges as well as the temperature and pressure conditions of the samples are summarised. The styrene oxide sample was purchased from Sigma Aldrich (>97\% purity) and was used without further purification.  
\begin{figure*}[ht]
\centering
\includegraphics[scale=0.36]{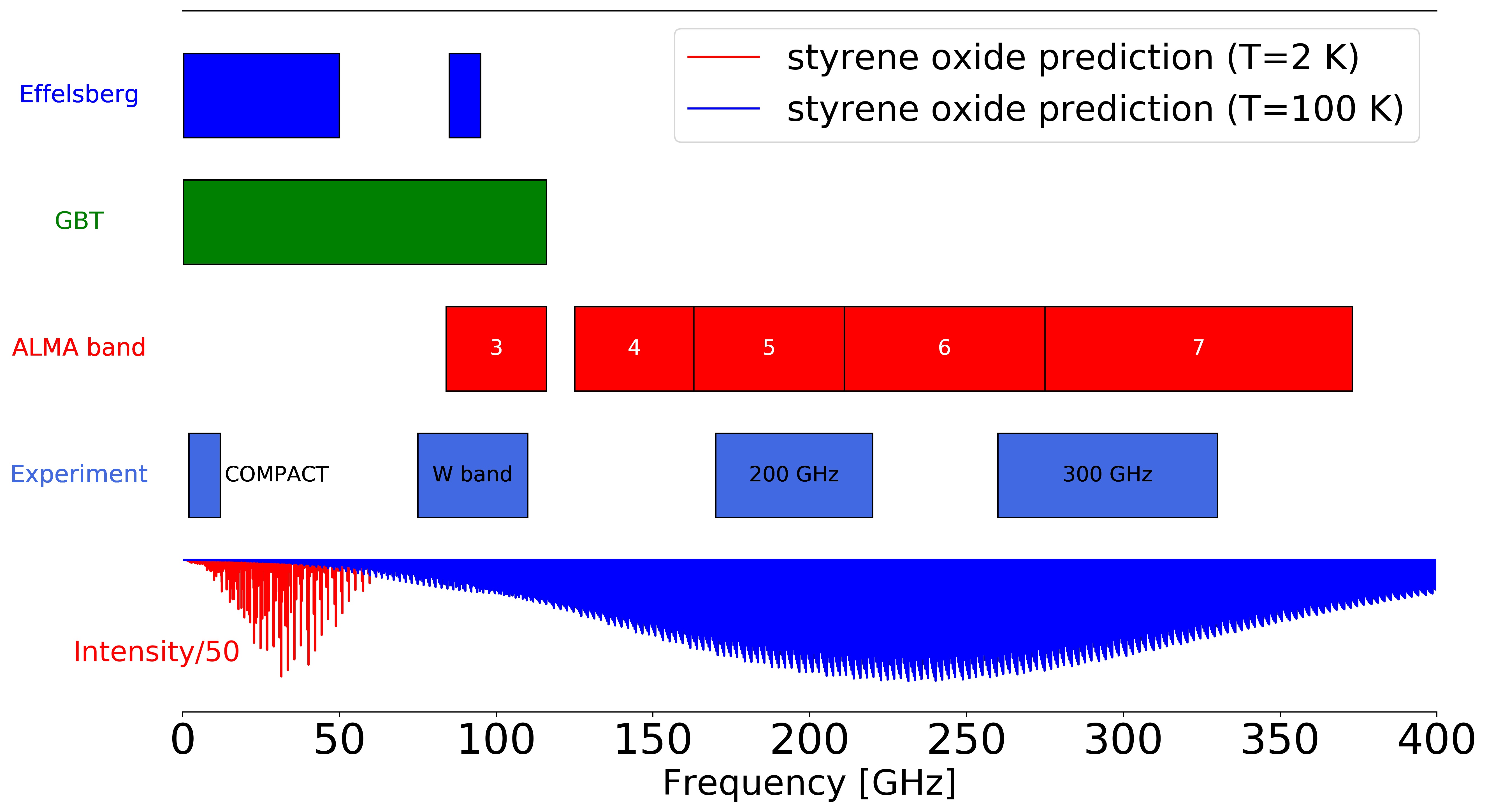}
\caption{Simulated overview spectrum of $\mathrm{C_{6}H_{5}C_{2}H_{3}O}$ in the millimeter and submillimeter wavelength region predicted at 2 K and 100 K. The intensity of the 2 K prediction has to be multiplied by a factor of 50 (bottom). Coverage of the Effelsberg Telescope, 100m Robert C. Byrd Green Bank Telescope (GBT) and ALMA bands   (top).}
\label{fig:bands}
\end{figure*}

\begin{center}
\begin{table*}[ht]
\centering
\caption{
Overview of the different spectrometers and sample conditions of styrene oxide measurements, including  the rotational temperature $T$, the operating pressure $p$, and full width at half maximum (FWHM) linewidth of recorded lines
%measured frequency ranges and experiments with setup, simulated rotational temperature $T$, typical pressure $P$, and linewidth (FWHM).
}
\begin{tabular}{ccccccc}
\hline 
Range /GHz & Figure&  Spectrometer / Method  &  Sample & $T$ /K &  $p$ /Pa & FWHM /kHz \\
  \hline
2-8  & \ref{meas:28} & COMPACT / CP-FT& jet & 1.5& $\mathrm{5\cdot 10^{-4}}$& 60\\ 
%\hline 
8-12  & \ref{meas:812} & COMPACT / CP-FT& jet& 2 & $\mathrm{5\cdot 10^{-4}}$&70\\ 
%\hline 
75-110  & \ref{meas:Wband} & W-band / CP-FT&flow cell& 335 & 1&500\\ 
%\hline 
%\hline
170-220& \ref{meas:200zoom} &Kassel THz / 2f& static cell& 293.15 & 5 &600\\ 
260-330 & \ref{meas:300zoom}&Kassel THz / 2f& static cell& 293.15&5& 600\\ 
\hline 
%\hline
\end{tabular} 
\label{tab:overview}
\end{table*}
\end{center}
The Hamburg COMPACT spectrometer, which operates between 2-18 GHz, is fully described elsewhere \cite{DavidSchmitz.2012,Perez.2016,Domingos.2017,Fatima.2020}. It was used to record the spectrum in the regions 2-8 GHz and 8-12 GHz. 
The sample was placed in the reservoir of a heatable nozzle and was supersonically expanded into the vacuum chamber with neon %at a pressure of  XXX 
as the carrier gas. During the 2-8 GHz experiment, the reservoir was heated to 60 $^{\circ}$C,  and it was heated to 100 $^{\circ}$C in the 8-12 GHz experiment. An arbitrary waveform generator (AWG) generates a train of eight chirped pulses, where each chirped pulse covers the whole bandwidth of the excitation frequency range, and which are amplified and transmitted into the vacuum chamber. For the 2-8 GHz experiment, the pulses are amplified by a 300 W travelling wave tube amplifier; for the 8-12 GHz experiment, a 50 W solid-state amplifier was used. The free induction decay (FID) of the macroscopic polarisation is then recorded as a function of time and fast Fourier transformed into the frequency domain.  For the 2-8 GHz and 8-12 GHz experiments, individual FIDs were recorded for 40 $\mu$s.
To achieve a good signal-to-noise ratio (SNR), a total of 3 million FIDs were co-added for the 2-8 GHz region and 1 million FIDs for the 8-12 GHz region. 
\\
\\
The BrightSpec W-band spectrometer is a segmented CP-FTmmw spectrometer probing a gas flow at room temperature conditions.
The spectrum was obtained in the 75-110 GHz region. The W-band spectrometer operates in the segmented fashion, i.e., the 35 GHz of bandwidth is split up into smaller segments, and the spectrum across the whole bandwidth is achieved by concatenating the segments together \citep{Neill:13,Arenas.2017}. An AWG produces excitation pulses between 1.56-2.28 GHz, which are then frequency up-converted, amplified, and frequency multiplied (by the action of a x6 active multiplier chain (AMC) with about 30 mW output power). The excitation chirp is coupled into the vacuum chamber via horn antennas, where the radiation interacts with the molecular sample, and the FID of the induced polarisation is collected and down-converted. The time domain signal is digitised using a real-time digitiser and fast Fourier transformed into the frequency domain. In the high dynamic range (HDR) mode, as was applied in this study, the spectrum is split into segments of 30 MHz bandwidth \citep{Arenas.2017}. For this experiment, 500 000 FIDs were co-added per segment, resulting in a measuring time of approximately 50 minutes. In contrast to the lower frequency measurements, the W-band experiment was run with the vacuum chamber acting as a room temperature slow flow cell. For this, the sample was placed in an external reservoir, pre-treated in a freeze-pump-thaw manner, and heated to 60 $^{\circ}$C. The vapour was allowed to flow into the vacuum chamber, where a pressure of approximately 1 Pa was maintained.
\\
\\
In the experiment with the Kassel THz spectrometer, a 2f frequency modulation technique was applied to probe the molecules in a static vacuum glass cell. Frequency modulation absorption spectra between 170 GHz and 220 GHz as well as between 260 GHz and 330 GHz were collected. The  Kassel THz spectrometer can also be used for studies up to 800 GHz (25.9 $\mathrm{cm^{-1}}$), as has been shown by Herberth \textit{et al.}  \cite{HERBERTH201937}. For the absorption experiments, the styrene oxide sample was measured in a static vacuum cell of 3 m length with 5 Pa total pressure. The sample was heated in a long-necked flat-bottomed glass flask until there was enough vapor pressure in the chamber. We used a \textit{Keysight 83650B} signal generator for 2f modulation. The 2f frequency deviation was set around 500 kHz having a modulation frequency of 47.115 kHz. The modulated microwave frequencies were transmitted to an AMC (x12, 30 mW output power) from \textit{Virginia Diodes Inc}. The up-converted signal was then focussed by Teflon lenses and broadcast into the glass cell. Standard gain horn antennas WR-4.3, WR-3.4, and WR-2.8 with a typical gain of 20 dBi transformed the waveguide beam into a free space wave and reverse. The absorption signal was collected by zero-biased room temperature detectors (from \textit{Virginia Diodes Inc.}). After passing a \textit{Stanford Research SR560} bandpass filter with 6 dB/octave, it was amplified and demodulated with a \textit{Signal Recovery 7280 DSP} lock-in-amplifier. The data processing was done using a LabVIEW program. Further data reduction, e.g. baseline substraction, was performed using a home-written python code. 
%\end{description}

\subsection{Quantum chemical calculations and spectral analysis} \label{calcs}
The computational chemistry program Gaussian 09 \citep{M.J.Frisch.2016b} was used to perform geometry optimisations and anharmonic frequency calculations (B3LYP/aug-cc-pVTZ, B3LYP/def2-TZVP, B3LYP/6-311++G(d,p), and MP2/6-311++G(d,p)). Structural parameters, rotational constants, and dipole moment components were retrieved. 
The calculated rotational constants, $A_{0}$, $B_{0}$, and $C_{0}$, and quartic and sextic corrections are given in Table \ref{tab:results} in Section \ref{res}. 
$A_{0}$, $B_{0}$, and $C_{0}$ are the centrifugal distortion corrected rotational constants. In the following the \textit{0}-suffix is omitted and the rotational constants are denoted by $A$, $B$, and $C$.  
With the equation $\kappa=\frac{2B-A-C}{A-C}$ from \citet{Townes.2013} and the calculated rotational constants, we derived Ray's asymmetry parameter $\kappa_{calc}\approx$-0.90. This means that styrene oxide has a near-prolate asymmetric rotor spectrum. 
In the rotational spectrum, all electric dipole transitions ($a$-type, $b$-type, and $c$-type) occur, resulting in a dense spectrum, especially at room temperature. The individual dipole moment components are obtained from the quantum chemical calculations (B3LYP/def2-TZVP) and are about $\mu_{a}$=0.6 D, $\mu_{b}$=1.0 D and $\mu_{c}$=1.3 D.
The overall dipole moment of SO was calculated to be $\mu_{total}$=1.8 D. The simulated rotational spectrum for the vibrational ground state up to $J_{max}$=120 and 330 GHz consists of $~12\cdot10^{4}$ rotational transitions. Anharmonic frequency calculations were carried out in order to determine the energies of the vibrationally excited states and their corresponding rotational constants. Styrene oxide has 3N - 6 = 45 normal modes, with N=17 being the number of atoms. With regard to a previous W-band spectroscopy experiment at room temperature for a molecule of similar size and a strong permanent dipole moment of 4 D \citep{Arenas.2017}, the instrument's sensitivity allows the detection of rotational transitions of vibrationally excited molecules that are about 400-500 $\mathrm{cm^{-1}}$ in energy above the vibrational ground state. Higher-lying vibrationally excited states (>500 $\mathrm{cm^{-1}}$) are weak in intensity, and at room temperature only the low-energy vibrations are sufficiently excited and considered in the analysis performed in this work.
\\
\\
We used PGOPHER \citep{ColinM.Western.2017} to assign experimental rotational lines of the vibrational ground state to our predictions. The jet measurements in the low frequency range were a good starting point for the analysis: firstly, the spectrum is simplified due to effective cooling, and secondly, at low frequencies, there are mostly rotational levels with low $J$ and $K$ quantum numbers. 
We continued the assignment including rotational levels with higher quantum numbers up to $J, K$=120, which correspond to rotational levels with higher transition frequencies, until the line assignment procedure was complete.
As a starting point for the assignment of the vibrationally excited states of SO, we first calculated the difference $\Delta$ between the experimental and predicted ground state rotational constants: $\Delta_{A}$, $\Delta_{B}$, $\Delta_{C}$=  $A_{exp}-A_{calc}$, $B_{exp}-B_{calc}$, and $C_{exp}-C_{calc}$. 
For the B3LYP/aug-cc-pVTZ and B3LYP/def2-TZVP anharmonic frequency calculations, the difference $\Delta$ of the constants from theory and experiment is given in Table S1 of the ESI\dag. We then accounted the theoretical rotational constants $A_{vib}$, $B_{vib}$, $C_{vib}$ from the anharmonic prediction of the vibrationally excited states for this difference $\Delta$ and obtained a new set of constants for all vibrational states: $A_{adj}=A_{vib}+\Delta_{A}$, $B_{adj}=B_{vib}+\Delta_{B}$, $C_{adj}=C_{vib}+\Delta_{C}$ (see Table S2 of the ESI\dag).
Additionally, all ground state centrifugal distortion constants up to sextic order were used as initial values for the fitting process. Due to spectral congestion and stronger centrifugal distortion effects, this was essential for further assignments. The adjustment of the constants $A, B, C$ to $A_{adj}$, $B_{adj}$, and $C_{adj}$ as described above with regard to the calculations is summarised in Table S2 of the ESI\dag.

\section{Experimental results and discussion} \label{res}
\subsection{Vibronic ground state of styrene oxide}
We used different high-resolution spectrometers to measure rotational spetra of SO in a frequency range from 2-330 GHz (see Table \ref{tab:overview}). 
Selected portions of the recorded styrene oxide spectra are shown in Figures \ref{meas:28}-\ref{meas:300zoom} for the different frequency regions. 
The upper traces (blue) show the recorded spectra, and the lower traces (red) show the simulated spectra, which are based on molecular parameters from a least-squares-fit data analysis (see Table \ref{tab:results}).  
For the labeling of the transitions we use the notation $\mathrm{\Delta K_{a}\Delta J_{K_{a}K_{c}(J)}}$ with $K_{a}, K_{c}, J$ denoting the quantum numbers of the lower rotational state, and $\mathrm{\Delta K_{a}}$ and $\mathrm{\Delta J}$ are respectively designated with p,q,r and P,Q,R, giving the change in quantum numbers $\mathrm{\Delta}$ = -1, 0, and +1. In cases where $\mathrm{\Delta K_{c}}$ needs to be specified we added the term \textit{b-} or \textit{c-type} according to $\mathrm{\Delta K_{c}} = \pm1$ or $\mathrm{\Delta K_{c}}$ = 0, respectively.
\\
\\
The broadband W-band measurement at room temperature exhibits a very dense spectrum with thousands of rotational lines. Many of them belong to low-lying vibrational modes. 
The jet measurements at low frequencies are less dense due to effective cooling of rotational and vibrational degrees of freedom. Typically, only the vibronic ground state is populated under jet conditions.
The linewidths of the recorded transitions vary from 60 kHz at low frequencies (2-8 GHz) and low temperature (1.5 K) to 600 kHz at high frequencies (260-330 GHz) and room temperature.
All three permanent electrical dipole components are distinct from zero and thus, $a$-, $b$-, and $c$-type transitions are observed throughout the whole frequency range up to 330 GHz. 
\\
\\
%{\textbf{2-8 GHz spectrum of styrene oxide:}
%\paragraph{\textbf{2-8 GHz spectrum of styrene oxide:}}
In Figure \ref{meas:28}, the recorded spectrum and a 1.5 K simulated spectrum based on parameters derived from a least-squares-fit analysis of SO from 2-8 GHz are shown. The strongest features in the spectrum arise from low $J$ and $K$ ($J,K$<5) transitions of Q- and R-branches. Up to 6.3 GHz, transitions of the $\mathrm{rQ_{0}}$ branch dominate the spectrum.
The typical linewidth (FWHM) is about 60 kHz and was determined using Gaussian fits.
Excitation of the molecules with an intense pulse of 300 W and averaging over 3 million individual spectra results in a SNR as high as 9500:1 for the strongest transitions.
%The SNR can be as high as 9500:1. 
In Figure \ref{meas:28}, an enlarged section shows a spectral excerpt at around 6.4 GHz with $a$-, $b$-, and $c$-type transitions of Q- and R-branches. 
\begin{figure*}[ht]
\centering
%[width=\hsize]
%\includegraphics[width=\hsize]{img/styrene_oxide_2_8_GHz_mV_abc2.png}
\includegraphics[scale=1.2]{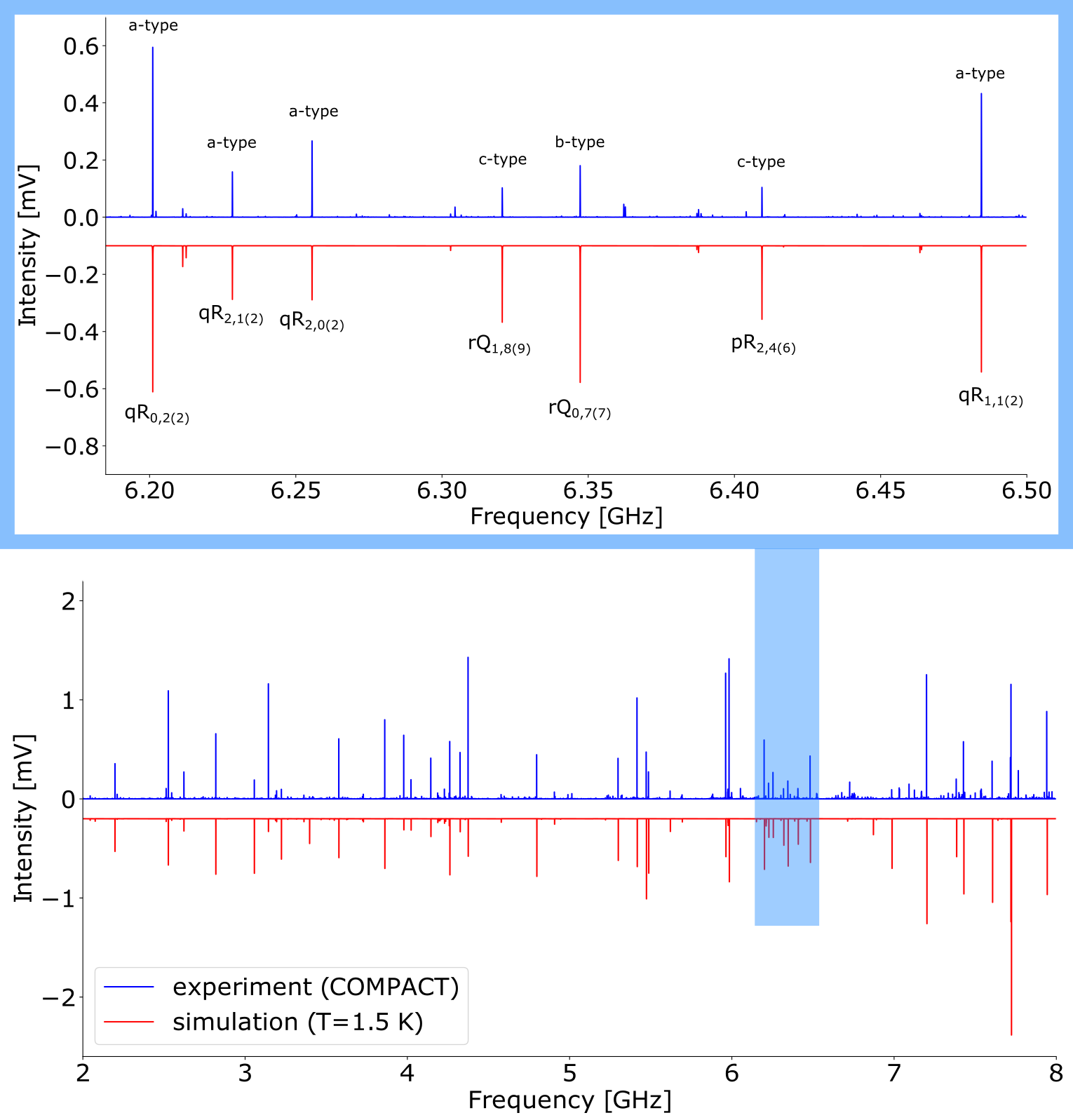}
\caption{Bottom graph: Spectrum of styrene oxide in an adiabatically cooled supersonic jet expansion at 1.5 K, measured with the Hamburg COMPACT spectrometer between 2-8 GHz. The upper trace (blue) shows the measured spectrum, the lower trace (red) shows the simulated spectrum based on molecular parameters derived from a least-squares fit analysis. Top graph: The enlarged spectral region shows examples of $a$-, $b$-, and $c$-type transitions at 6.35 GHz (top).
}
\label{meas:28}
\end{figure*}
\\
\\
%\textbf{8-12 GHz spectrum of styrene oxide:} 
%\paragraph{\textbf{8-12 GHz spectrum of styrene oxide:}} 
In Figure \ref{meas:812}, the recorded spectrum and a simulated spectrum of adiabatically cooled SO at 2 K from 8-12 GHz is shown. Analyzing the data reveals a slightly higher rotational temperature of 2 K in the molecular jet, compared to the 2-8 GHz measurements. Throughout the range from 6.3 GHz up to 12 GHz, $\mathrm{rQ_{1}}$-branch and R-branch transitions dominate the spectrum. The strongest signal measured at 8236.76 MHz has a SNR of 3725:1 and a linewidth of 70 kHz, which is a typical linewidth for the COMPACT measurements in this region. 
The overall higher SNR of the 2-8 GHz experiment compared to the 8-12 GHz experiment is explained by the number of averages (3 million and 1 million, respectively) and the output power of the amplifiers (300 W and 50 W, respectively).
\begin{figure*}[ht]
\centering
\includegraphics[width=\hsize]{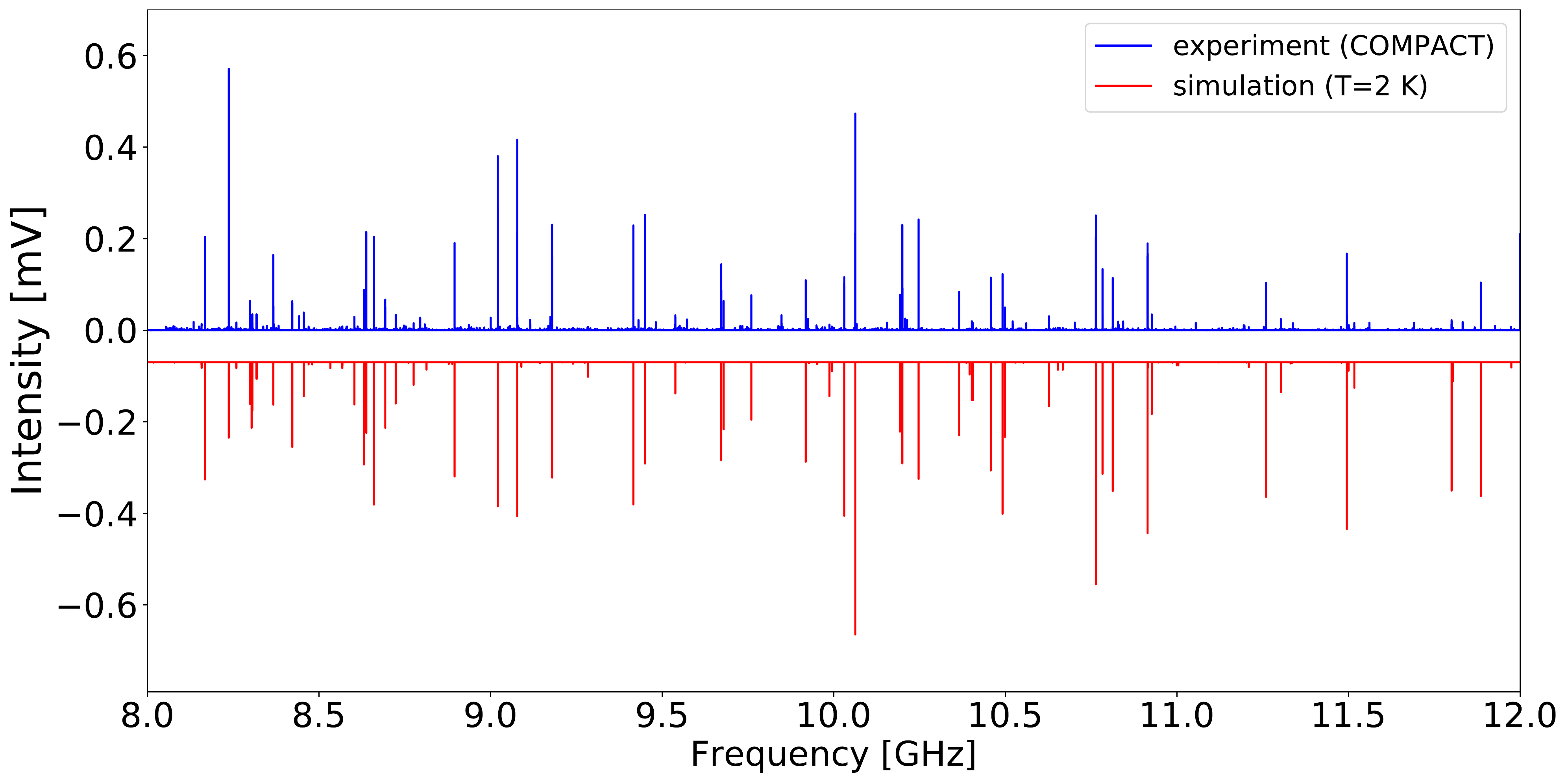}
\caption{
Spectrum of adiabatically cooled styrene oxide between 8-12 GHz measured with the Hamburg COMPACT spectrometer in a supersonic jet (upper trace, blue) and the simulated spectrum (T=2 K) based on parameters derived from a least-squares-fit analysis (red).
}
\label{meas:812}
\end{figure*}
\\
\\
The recorded W-band spectrum from 75-110 GHz and a simulated spectrum at a rotational temperature of 335~K is shown in Figure \ref{meas:Wband}.
This is approximately the temperature the sample reservoir was kept at during the experiment.
The higher temperature, compared to the previously presented spectra, results in a dense spectrum of prominent $\mathrm{rQ_{K_{a}}}$ branches ($K_{a}$ = 11-16), separated by roughly 
%\begin{eqnarray}
$2\cdot(A-\dfrac{B+C}{2}) \approx 6.6~\mathrm{GHz}$,
%\end{eqnarray}
which can be seen in Figure \ref{meas:Wband} ($A$, $B$, and $C$ are taken from Table \ref{tab:results}). In addition, series of equidistant $\mathrm{R_{K_{a}}}$-branch transitions dominate the spectrum, with the strongest lines originating from low $\mathrm{K_{a}}$ levels. 
In Figure \ref{meas:Wband}, a zoom-in of the spectrum around 86 GHz is included. Several assigned lines are  labeled by their quantum numbers. 
As can be seen from the enlarged spectral region in Figure \ref{meas:Wband}, the simulated spectrum, based on the fitted rotational parameters, accurately matches the measured spectrum with regard to the frequency positions and relative intensities, including blended lines, e.g., between $\mathrm{rR_{9,1(10)}}$, $\mathrm{rR_{9,2(10)}}$ and $\mathrm{rR_{14,64(77)}}$, $\mathrm{rQ_{13,49(61)}}$. The energy level diagram of the $\mathrm{Q_{13,48(61)}}$ doublet shows both $b$- and $c$-type transitions (see zoom-in, Figure \ref{meas:Wband}). %Note the stronger $c$-type transition.
Due to increasing Doppler broadening at higher frequencies and temperatures, the W-band transitions of SO have typical linewidths of 500 kHz. The SNR was observed up to 100:1 for the strongest transitions. 
\begin{figure*}[ht]
\centering
\includegraphics[scale=.65]{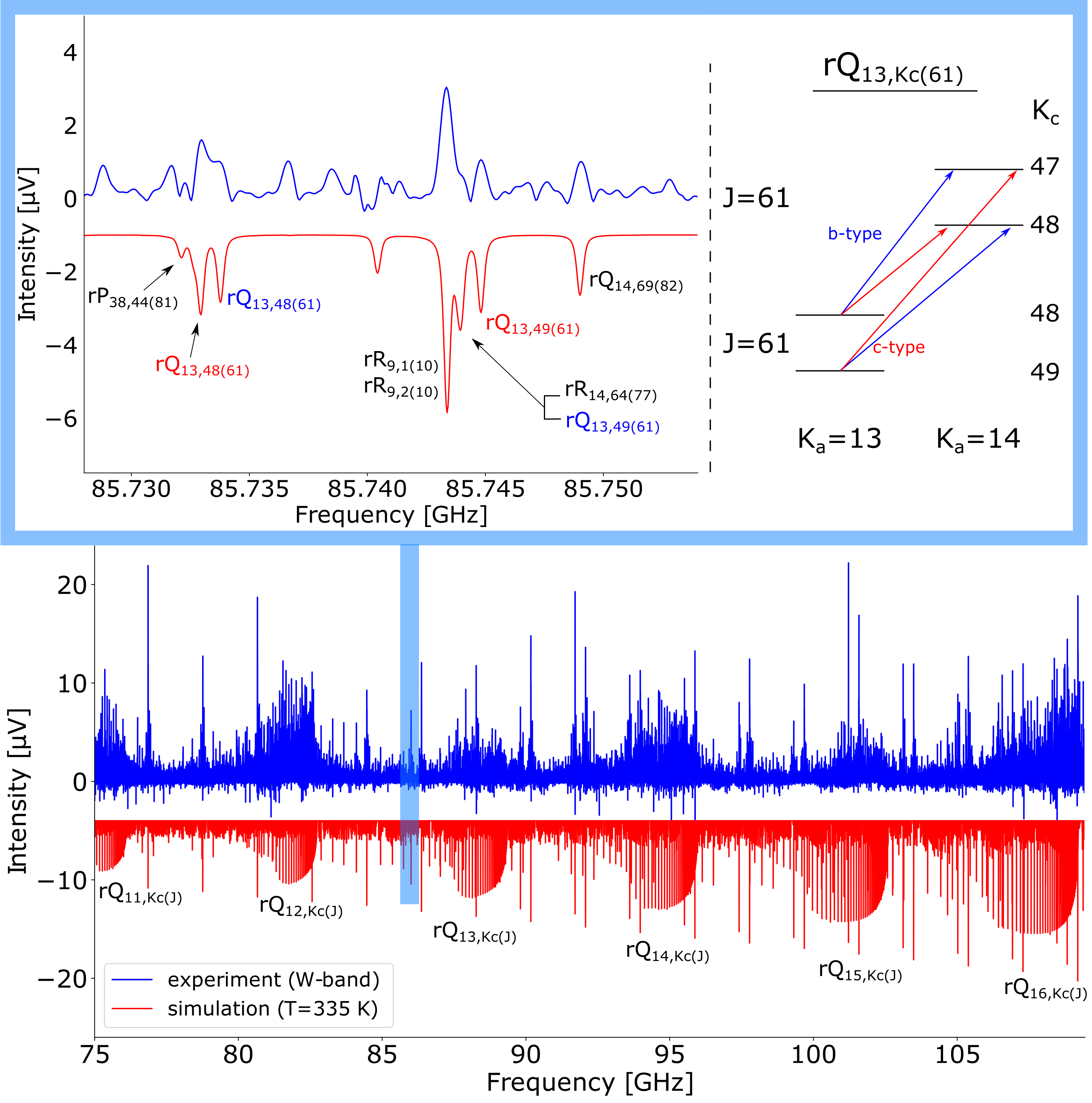}
\caption{Bottom graph: Spectrum of styrene oxide between 75-110 GHz recorded with the W-band spectrometer in a gas flow kept at a temperature of about 335 K (upper trace, blue) and simulated spectrum (lower trace, red) at 335 K (bottom). Top graph: Zoom-in around 85.74 GHz with the measured spectrum (upper trace, blue) and the simulated spectrum (lower trace, red) with quantum numbers assigned to the transitions (left top). Schematic energy level diagram of the $\mathrm{rQ_{13,K_{c}(61)}}$ doublet with $b$- and $c$-type transitions (right top).
}
\label{meas:Wband}
\end{figure*}
\\
\\
%\textbf{170-220 GHz spectrum of styrene oxide:}
%\paragraph{\textbf{170-220 GHz spectrum of styrene oxide:}}
Spectra between 170-220 GHz and 260-330 GHz were recorded with the Kassel THz spectrometer in 2f modulation (see Table \ref{tab:overview}). As an example, Figure \ref{meas:200zoom} shows a narrow spectral portion, which encompasses the transitions $J$ = 65, and 66 of the $\mathrm{rQ_{26}}$-branch and two $\mathrm{rR_{K_{a}}}$-branch transitions together with the simulated room temperature spectrum. Each line consists of two pairs of unresolved $b$- and $c$-type transitions. The SNR is lower than that of the CP-FT spectra, and therefore weak transitions like the \textit{a}-type doublet $\mathrm{qR_{40,K_{c}(83)}}$ on the left side of the spectrum were excluded from the fit.
\begin{figure*}[ht]
\centering
\includegraphics[scale=.38]{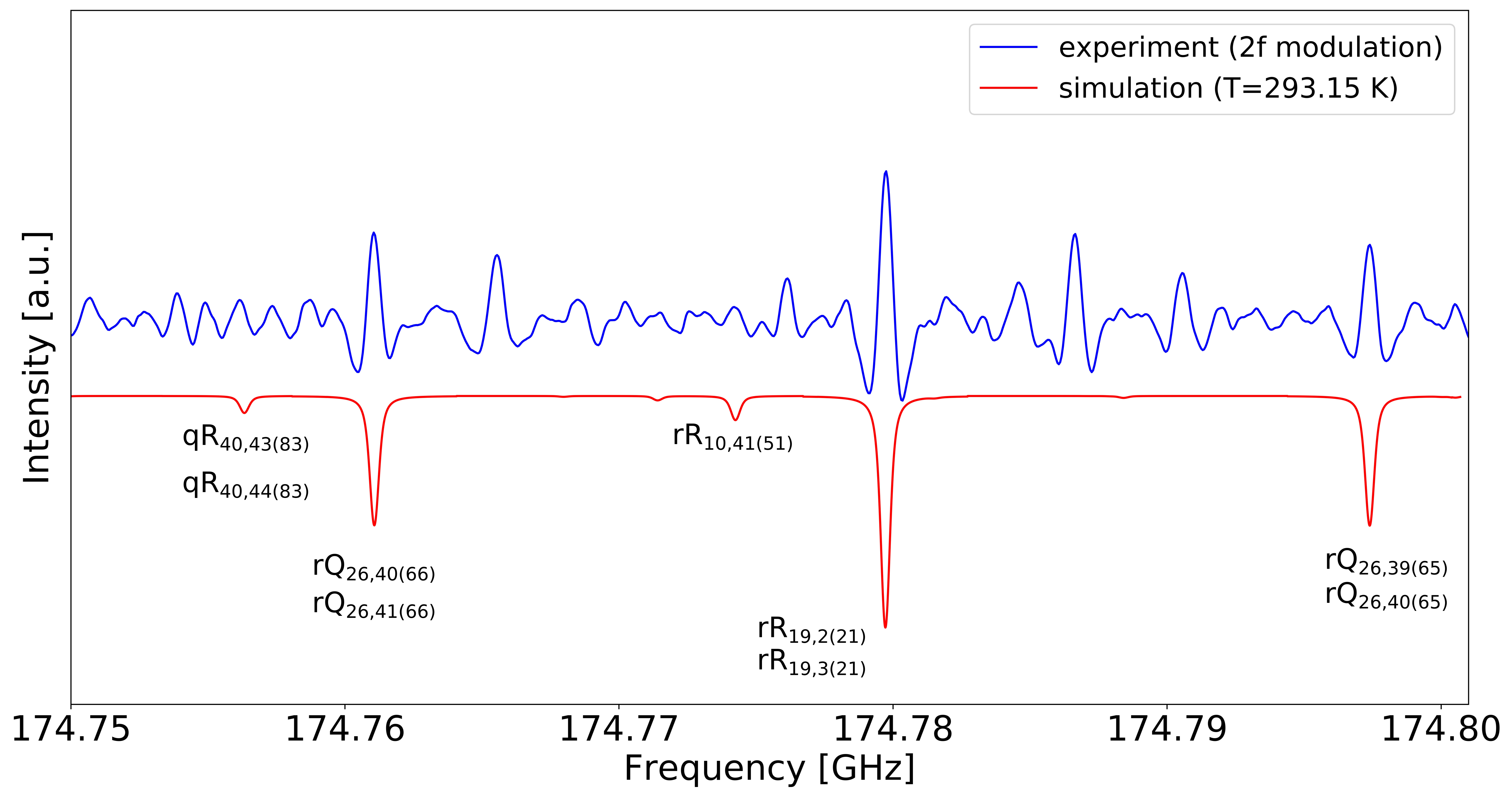}
\caption{Spectral excerpt between 174.75-174.80 GHz of styrene oxide at room temperature (upper trace, blue) with the Kassel THz experiment and simulated spectrum (lower trace, red) based on parameters derived from a least-squares-fit analysis. The excerpt displays the 2f modulation signals of the $\mathrm{rQ_{26}}$-branch.} % $\mathrm{rQ_{26,K_{c}(J)}}$ from $J$=27 to higher $J$ values
\label{meas:200zoom}
\end{figure*}
%\paragraph{\textbf{260-330 GHz spectrum of styrene oxide:}}
\\
\\
%\textbf{260-330 GHz spectrum of styrene oxide:} 
Figure \ref{meas:300zoom} shows the $\mathrm{rQ_{43}}$-branch transitions $J$=44 to $J$=111 and several strong $\mathrm{rR_{K_{a}}}$ transitions around 287 GHz. 
In addition, several lines were found in the spectrum that could not be assigned to the ground state of SO, but during the data analysis it became apparent that these additional lines belong to low-lying vibrational modes, thermally excited at room temperature.
This will be discussed in more detail in Subsection \ref{Vibes}. 
\begin{figure*}[ht]
\centering
\includegraphics[width=\hsize]{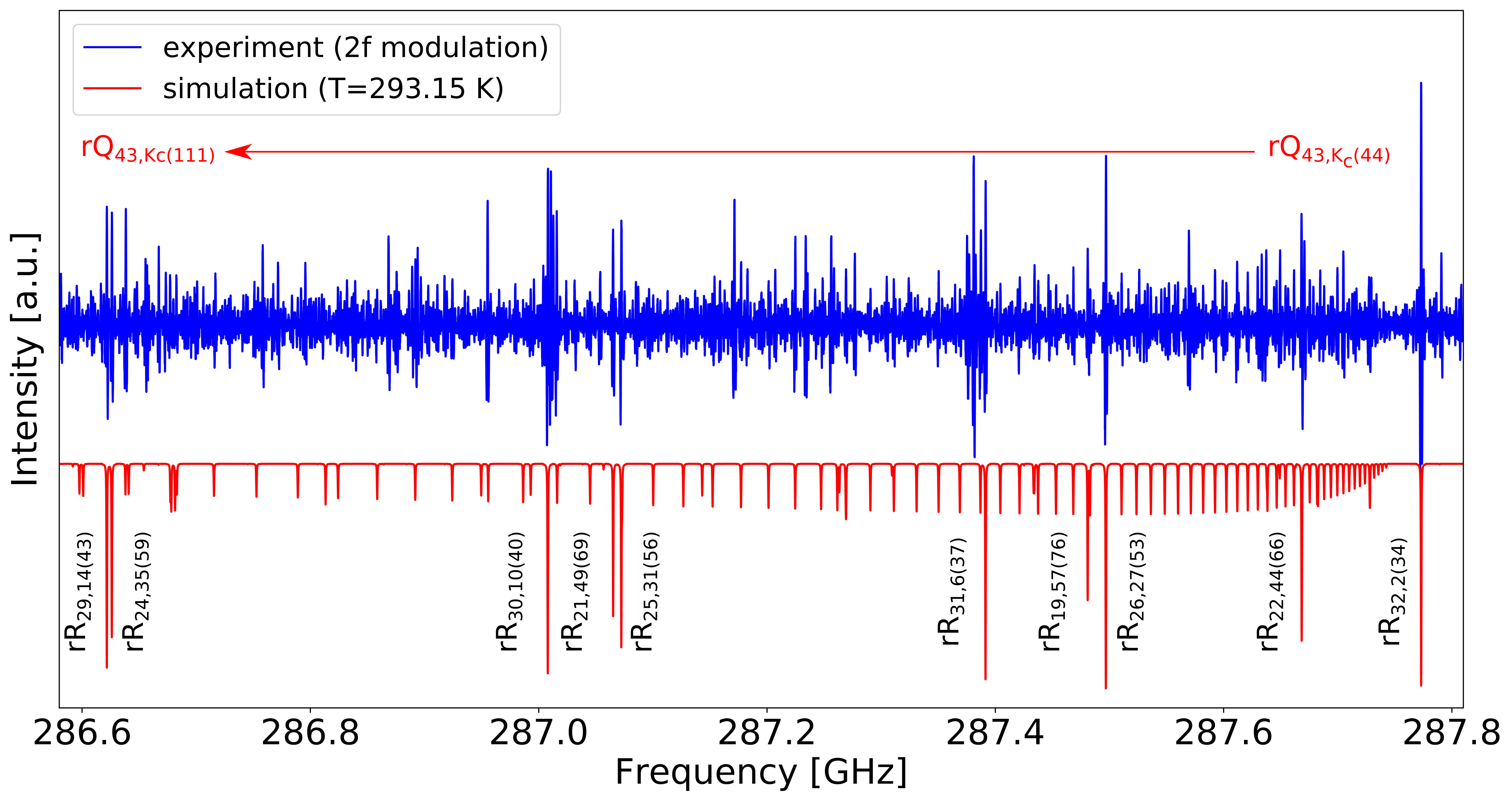}
\caption{Spectral portion between 286.6-287.8 GHz of styrene oxide at room temperature with the Kassel THz absorption experiment (upper trace, blue) and simulated spectrum (lower trace, red) with quantum number labels based on parameters derived from a least-squares-fit analysis. The excerpt displays the 2f modulation signals of the $\mathrm{rQ_{43}}$-branch from $J$=44 to $J$=111.}
\label{meas:300zoom}
\end{figure*}
The linewidths of transitions measured in the 2f modulation range from 500 kHz to 700 kHz, which is due to a five times higher pressure compared to the W-band measurements, and due to the 2f frequency deviation, which provides the strongest signals when set to a value half the line width. 
\\
\\
%\textbf{Vibronic ground state results:}
The results from our experiments between 2 GHz and 330 GHz, and the quantum chemical calculations are summarised in Table \ref{tab:results}.
The asymmetric top Hamiltonian in Watson's A-reduction was used in the \textit{Ir} representation for this molecule with $C_{1}$ symmetry. 
We assigned transitions up to $J$=120 and up to $K_{a}=49$ as well as $K_{c}=103$. Experimental rotational constants and %and the quartic as well as sextic 
centrifugal distortion constants up to sextic order have been determined and are given in Table \ref{tab:results}. In total, 6603 ground state transitions from all frequency regions are included in the global fit. The total average uncertainty (1$\mathrm{\sigma}$ confidence) is below 54 kHz. The largest contributions to the uncertainty result from high $J$ and $K$ transitions, particularly at large differences between $K_{a}$ and $K_{c}$.
The quantum chemical calculations are in good agreement with the experimentally derived values, especially the calculated Ray's asymmetry parameter ($\kappa_{calc}\approx$-0.8996), which deviates by only about 0.2 \% from the experimental result of  $\kappa_{exp}$=-0.8977.
\\
\\
In the following, the calculated centrifugal distortion corrected rotational constants are compared to experiment.
The calculated rotational constants $A$, $B$, and $C$ agree within 1.6 \% with the experimental values, with largest discrepancies being found for the 6-311++G(d,p) basis set used for both DFT B3LYP and \textit{ab initio} MP2 calculations. 
Here, the $A$ constant in the B3LYP calculation is closer to the experimental value than in the MP2 calculation (0.7 \% compared to 1.5 \% relative deviation), but for the constants $B$ and $C$, it is the reverse (0.9 \% and 1.2 \% compared to 1.6 \% and 1.5 \%).
For the B3LYP/aug-cc-pVTZ and the B3LYP/def2-TZVP calculations, the deviation from the experimental values of $A$, $B$, and $C$ are comparable and about 0.5 \% for all primary constants. 
We found that the calculations performed here underestimate the $B$ and $C$ rotational constants compared to experiment. 
The B3LYP/aug-cc-pVTZ and B3LYP/def2-TZVP calculations agree best with the experimental results.
The calculated quartic centrifugal distortion constants are in good agreement with the experimental parameters. The sextic centrifugal distortion constants show larger deviations, however these values were not used at the beginning of the line assignment process. Experimental values for the sextic constants were eventually derived using specifically high $J$, $K_{a}$, and $K_{c}$ transitions.
Overall, all quantum chemical methods used here reach good agreement with the experiment.
\begin{center}
\begin{table*}[ht]
\centering
\caption{Rotational constants and quartic and sextic centrifugal distortion constants for the ground state of styrene oxide with uncertainties described in Watson's A-reduction and predicted rotational parameters obtained from anharmonic frequency calculations. The frequency uncertainty $\sigma$ is the single standard deviation from the fit with N=6603 lines being included. The Ray's asymmetry parameter $\kappa$ is shown as derived from the experiment and at different levels of theory
}
\begin{tabular}{p{1.7cm}ccccc}
\hline 
   & experiment & B3LYP		&	B3LYP	& B3LYP	 		& MP2 \\
  & 	& aug-cc-pVTZ & def2-TZVP & 6-311++G(d,p) & 6-311++G(d,p) \\
\hline 
%\hline  
$A$ /MHz &\tablenum{4348.864334(75)} & \tablenum{4361.78} &\tablenum{ 4359.15 }& \tablenum{4319.88}& \tablenum{4283.84}\\ 
%\hline 
$B$ /MHz & \tablenum{1124.882753(39)} & \tablenum{1116.87 }&\tablenum{ 1117.78}& \tablenum{1107.26 }&\tablenum{ 1114.38}\\ 
%\hline 
$C$ /MHz & \tablenum{951.226312(45)} & \tablenum{945.91} & \tablenum{946.51}& \tablenum{936.87} & \tablenum{940.11}\\ 
%\hline 
%\hline
$\Delta_{J}$ /kHz & \tablenum{0.0457603(83)} &\tablenum{0.045}  & \tablenum{0.045}& \tablenum{0.045 }& \tablenum{0.046} \\ 
$\Delta_{JK}$ /kHz & \tablenum{0.136478(28) }& \tablenum{0.14} & \tablenum{0.17} &\tablenum{0.16} & \tablenum{0.14} \\ 
$\Delta_{K}$ /kHz &\tablenum{ 0.512624(62)} &\tablenum{0.51} & \tablenum{0.50}&\tablenum{0.53 }& \tablenum{0.54}\\ 

$\delta_{J}$ /kHz& \tablenum{0.0078679(15)} &\tablenum{0.0078} & \tablenum{ 0.0077 }&\tablenum{0.0076 }&\tablenum{ 0.0077}\\ 
$\delta_{K}$ /kHz &\tablenum{ -0.52176(10)} &\tablenum{ -0.58} & \tablenum{-0.68}&\tablenum{-0.54}& \tablenum{-0.44}\\ 
$\Phi_{J}$ /mHz & \tablenum{0.00194(50)}& \tablenum{  0.003} & \tablenum{0.003}&\tablenum{0.003} &   \tablenum{  0.004}\\ 
$\Phi_{JK}$ /mHz &\tablenum{ -0.4611(18)}& \tablenum{-0.65 }&\tablenum{ -0.92}&\tablenum{-0.93}& \tablenum{ -0.46}\\ 
$\Phi_{KJ}$ /mHz &\tablenum{ -2.1599(56)}& \tablenum{ 0.14 }& \tablenum{0.07 }&\tablenum{-0.61 }&\tablenum{ -0.18}\\ 
$\Phi_{K}$ /mHz &\tablenum{ 1.674(17)} &  \tablenum{ -2.89 }& \tablenum{  -2.90}&\tablenum{0.88}& \tablenum{ -3.13}\\ 
$\kappa$ & \tablenum{ -0.8977 } & \tablenum{-0.8999}& \tablenum{-0.8996}&\tablenum{-0.8993} & \tablenum{-0.8957}\\
%\hline
%\hline
$\sigma$ /kHz & \tablenum{53.8 }& \\ 
%\hline
No. of lines& 6603 & \\%\multicolumn{2}{c}{number of transitions }\\ 
\hline 
%\hline
\end{tabular} 
\label{tab:results}
\end{table*}
\end{center}

\subsection{Vibrationally excited states of styrene oxide} \label{Vibes}
Aside from ground state transitions, we found many additional lines from vibrationally excited states of SO, which were assigned to the three energetically lowest normal modes $v_{45}$, $v_{44}$, and $v_{43}$ that are thermally populated at room temperature. The lowest-energy normal mode, $v_{45}$, corresponds to the torsion of the oxirane group relative to the phenyl group. The second lowest, $v_{44}$, is the tilting of the oxirane group relative to the phenyl plane. The third lowest, $v_{43}$, is the back-and-forth-bending of the oxirane group in the phenyl plane. We were able to determine their rotational constants $A$, $B$, $C$, and centrifugal distortion constants up to the sextic order. Furthermore, we calculated the harmonic and anharmonic energies with respect to the vibrational ground state on the B3LYP/aug-cc-pVTZ and B3LYP/def2-TZVP levels of theory (see Tables \ref{tab:resultsvib} and S3 of the ESI\dag). 
\\
For the analysis, an asymmetric top Hamiltonian in Watson's A-reduction was used in the \textit{Ir} representation, and levels up to $J$=120 were included.
\begin{figure*}[ht]
\centering
\includegraphics[scale=0.75]{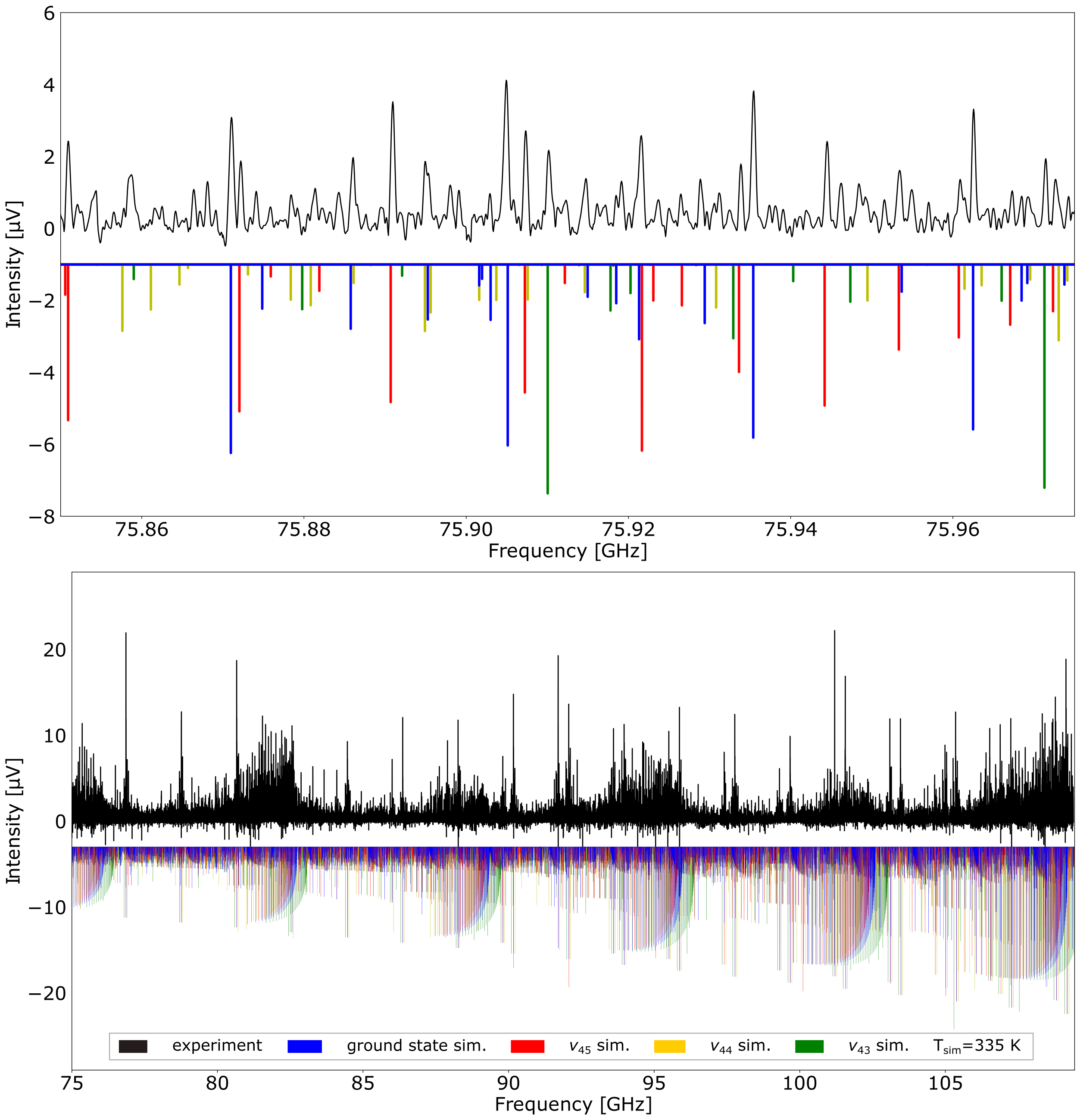}
\caption{Bottom graph: Spectrum of styrene oxide, kept at a temperature of about 335 K, between 75-110 GHz recorded with the W-band spectrometer in a gas flow (upper trace) and simulated rotational stick spectrum of the ground state and three excited states (lower traces) at 335 K. Top graph: The enlarged excerpt around 76 GHz shows the dense spectrum with the ground state prediction (blue) and the prediction of the vibrationally excited states (red, yellow, green) at 335~K.
}
\label{fig:vibW}
\end{figure*}
\begin{figure*}[ht]
\centering
\includegraphics[scale=0.63]{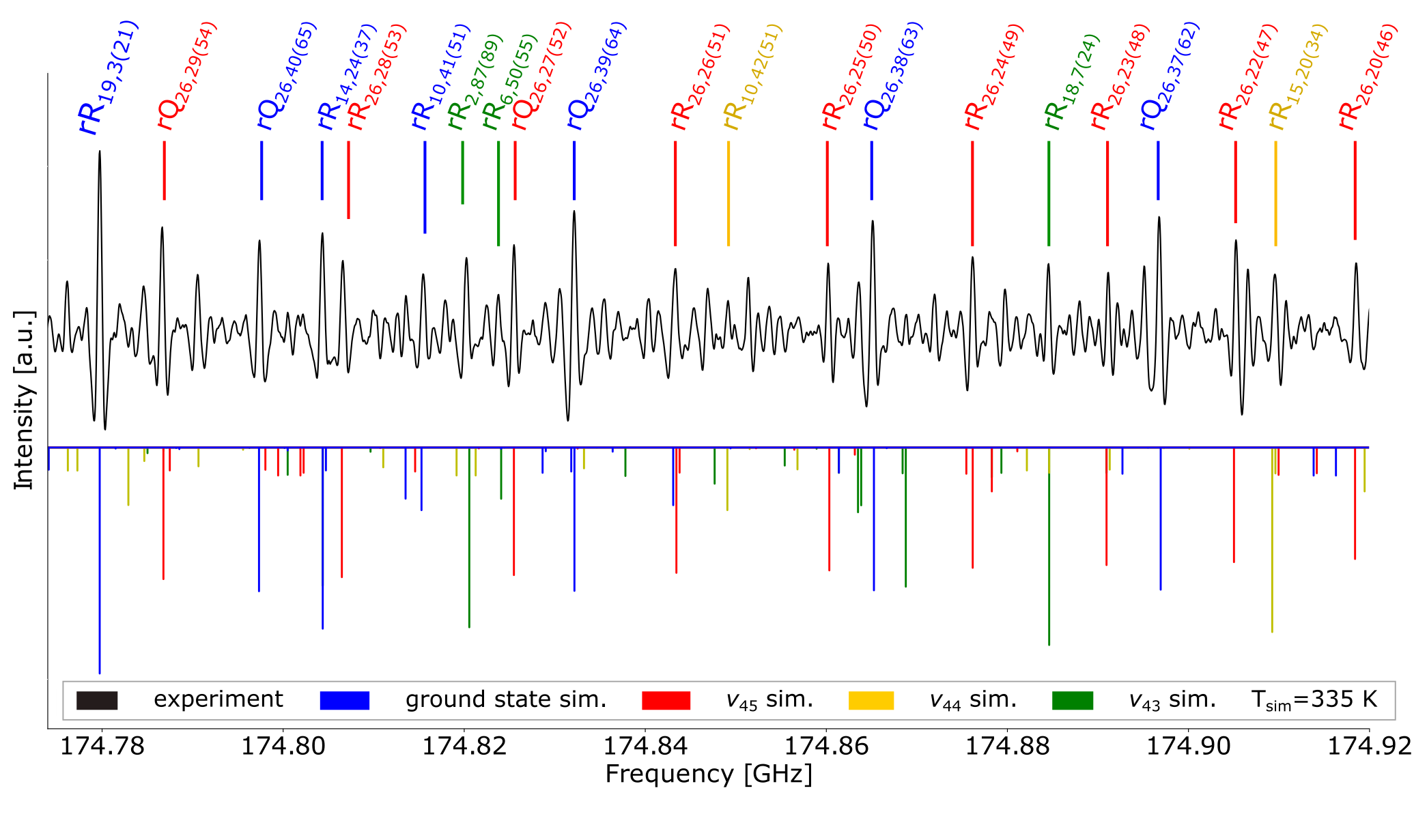}
\caption{
%Spectral excerpt between 174.78-174.92 GHz of styrene oxide 
Detailed excerpt of the styrene oxide spectrum at 174.85 GHz recorded at room temperature (upper trace) using the Kassel THz experiment and simulated rotational stick spectra of the ground state and three vibrationally excited states (lower traces) based on parameters derived from a least-squares-fit analysis (see Table \ref{tab:resultsvib}).
}
\label{fig:vib200}
\end{figure*}
In Figure \ref{fig:vibW}, the measured 75-110 GHz spectrum of SO is shown (upper trace) with the simulated rotational spectra of the ground state and three vibrationally excited states (lower traces). To enhance visibility, the simulated rotational spectra of the vibrationally excited states have been scaled to the same intensity as the ground state transitions. The enlarged spectrum in Figure \ref{fig:vibW} shows the line density and the decrease of intensity for the vibrationally excited states.  
Because of similar rotational constants in the ground state and in the assigned vibrationally excited states, the spectra strongly overlap, with dense Q-branches separated by approximately 6.6 GHz. In Figure \ref{fig:vib200}, a narrow spectral region at around 175 GHz is shown, presenting rotational transitions from the ground state and all three vibrationally excited states $v_{45}$, $v_{44}$, and $v_{43}$. 
The $\mathrm{rQ_{26}}$-branch transitions of the ground state and the lowest excited normal mode $v_{45}$ as well as strong R-branch transitions of $v_{44}$ and $v_{43}$ can be seen. The observed line intensity of the ground state is higher than those of the vibrationally excited states, and, as expected, the intensity decreases with increasing energy relative to the ground state. 
At frequencies above 300 GHz, only R-branch transitions and strong Q-branch transitions were assigned. The high frequency Q-branch assignments of the $v_{45}$ and $v_{43}$ states contributed the most to the frequency uncertainty of our fits.    
Derived rotational parameters for the vibrational modes $v_{45}$, $v_{44}$, and $v_{43}$ are summarised in Table \ref{tab:resultsvib}. We started the assignment of the vibrationally excited rotational transitions of SO using adjusted calculated rotational constants, as described in Subsection \ref{calcs} and summarised in Tables S1 and S2 of the ESI\dag. A total of 3000 to 4000 transitions were included in each fit. The average uncertainty is 158.6 kHz for $v_{45}$, 94.7 kHz for $v_{44}$, and 150.2 kHz for the $v_{43}$ state. Both the B3LYP/aug-cc-pVTZ and the B3LYP/def2-TZVP calculations, using the adjusted ground state constants (see Tables 3, and S1 and S2 of the ESI\dag),  agree well with the experimental results in Table \ref{tab:resultsvib}. Further lines can be seen in the spectra of styrene oxide that have not been assigned in our analysis and most likely origin from energetically low-lying vibrational hot bands and combination bands. Due to line confusion, line blending, and lower intensity, an accurate assignment was not realised.

%%%%%%%%%%%%%%%%%%%%%%%%%%%%%%%%%%%%%%%%%%%%%%%%%%%%%%%%%%%%%%%%%%%%%%%%%%%%%%%%%%%%%%%%%%%%%%%%%%%%
\begin{center}
\begin{table*}[ht]
\centering
\caption{Rotational constants for the three lowest vibrational modes $v_{45}$, $v_{44}$, and $v_{43}$ of styrene oxide with 1$\sigma$ uncertainties in the Watson's A-reduction. The calculated harmonic and anharmonic energies of the respective vibrational states based on the B3LYP/def2-TZVP level of theory together with the adjusted rotational constants are given as well
}
\begin{tabular}{lcccccc}
\hline 
  & \multicolumn{2}{c}{$v_{45}$}  & \multicolumn{2}{c}{$v_{44}$} & \multicolumn{2}{c}{$v_{43}$} \\
%\hline
& exp. & theory $^a$ & exp. & theory $^a$ & exp. & theory $^a$ \\
\hline
$E_{harm}$ /$\mathrm{cm^{-1}}$  &  &  \tablenum{55.05} &  & \tablenum{145.26} &  & \tablenum{195.20} \\
$E_{anharm}$ /$\mathrm{cm^{-1}}$  &  & \tablenum{71.93}  &  &  \tablenum{152.89} &  &  \tablenum{195.38} \\

$A$ /MHz & \tablenum{4342.87066(24)} &\tablenum{4344.83} & \tablenum{4336.83651(15)}& \tablenum{ 4335.68}& \tablenum{4362.57904(37)} & \tablenum{4366.56}
  \\ 
%\hline 
$B$ /MHz & \tablenum{ 1124.66433(39)}&\tablenum{1124.39}&\tablenum{1125.28101(12)  } & \tablenum{1125.50} &\tablenum{1124.14123(22) }& \tablenum{1124.10}\\ 
%\hline 
$C$ /MHz & \tablenum{ 951.27690(43)} &\tablenum{951.10}&\tablenum{951.78289(13)}& \tablenum{951.94} &\tablenum{950.42418(25) }&\tablenum{950.32} \\ 
%\hline
%\hline 
$\Delta_{J}$ /kHz & \tablenum{ 0.050632(47)} && \tablenum{0.045952(18)}  && \tablenum{0.045660(26)}&\\ 
$\Delta_{JK}$ /kHz & \tablenum{ 0.14142(13)} && \tablenum{0.151568(69)} && \tablenum{0.10961(17)  }& \\ 
%\hline 
$\Delta_{K}$ /kHz & \tablenum{ 0.51854(25)}&&\tablenum{0.35290(19)}  && \tablenum{0.87289(55)}& \\ 
%\hline 
$\delta_{J}$ /kHz & \tablenum{0.004472(15) }&& \tablenum{0.0079116(44)} && \tablenum{0.0078840(76)}&   \\ 
$\delta_{K}$ /kHz & \tablenum{ -0.6182(11) } && \tablenum{-0.52139(43)} && \tablenum{-0.50971(47)}&\\ 
%\hline 
%\hline 
$\Phi_{J}$ /mHz & \tablenum{0.4119(28) }&&\tablenum{ 0.00379(94)} && \tablenum{-0.0028(14) }& \\ 
%\hline 
%\hline 
$\Phi_{JK}$ /mHz &  \tablenum{0.0180(87)} && \tablenum{-0.4826(42) }&& \tablenum{-0.200(14)}&  \\ 
%\hline 
$\Phi_{KJ}$ /mHz & \tablenum{-5.962(37)} && \tablenum{-1.584(21) } & & \tablenum{-4.906(87)}& \\ 
$\Phi_{K}$ /mHz  & \tablenum{ 6.637(90) }&& \tablenum{-1.175(79)} && \tablenum{-3.45(23) }&  \\ 
%\hline
%\hline
%$\kappa_{exp}$ &  -0.8977  & -0.9009 \\
%\hline
%\hline
$\sigma$ /kHz &  \tablenum{158.6 }&& \tablenum{94.7 }&& \tablenum{150.2} &\\ 
%\hline
No. of lines&  \tablenum{ 3926} && \tablenum{3553} & &\tablenum{ 2992}&\\ 

\hline 
%\hline
\end{tabular} 
 $^a$ The theoretical $A, B,$ and $C$ values tabulated are the experimentally corrected $A_{adj}$, $B_{adj}$, and $C_{adj}$ of the B3LYP/def2-TZVP calculations
\label{tab:resultsvib}
\end{table*}
\end{center}
%%%%%%%%%%%%%%%%%%%%%%%%%%%%%%%%%%%%%%%%%%%%%%%%%%%%%%%%%%%%%%%%%%%%%%%%%%%%%

\subsection{$\mathrm{^{13}C}$ and $\mathrm{^{18}O}$ isotopologues of styrene oxide} \label{isotop}
Transitions of all singly-substituted $\mathrm{^{13}C}$ and $\mathrm{^{18}O}$ isotopologues of styrene oxide have been assigned in the 2-12 GHz region. 

\begin{table*}[ht]
\centering
\caption{Experimentally determined rotational constants $A$, $B$, $C$ for the ground state of the singly-substituted $\mathrm{^{13}C}$ and $\mathrm{^{18}O}$ isotopologues of $\mathrm{C_{6}H_{5}C_{2}H_{3}O}$. 
The parent species centrifugal distortion constants were kept fixed in the fitting process of the isotopologues (see Table \ref{tab:results})
}
\begin{small}
\begin{tabular}{cccccc}
\hline 
&$\mathrm{^{13}C}$-1&$\mathrm{^{13}C}$-2&$\mathrm{^{13}C}$-3&$\mathrm{^{13}C}$-4&$\mathrm{^{13}C}$-5\\
%\hline
\hline
$A$ /MHz&  
\tablenum{ 4286.4956(11)    }

& \tablenum{  4311.8212(14)}

 &    \tablenum{   4345.0365(14) }

&   \tablenum{4282.7224(12) }

&  \tablenum{ 4307.8539(15)  }
 \\ 
$B$ /MHz&
  \tablenum{ 1123.27646(35)  }
 &
  \tablenum{1113.28501(41) }
&
   \tablenum{   1108.39144(37)}
&
\tablenum{  1118.13792(34)}
&
\tablenum{ 1124.69685(41) }
\\ 
$C$ /MHz&   
 \tablenum{ 947.07189(28)  }
 &
   \tablenum{ 941.16512(31)   } 
    &
     \tablenum{ 939.25297(36)  }
&
 \tablenum{  943.22671(28)}
&
\tablenum{ 949.12352(34)  }
 \\ 
$\sigma$ /kHz  & \tablenum{	22.0 }	&\tablenum{	18.4	}&	\tablenum{16.1	}&	\tablenum{22.2}	&	\tablenum{24.2}	\\ 
No. of lines & \tablenum{	78} 	&\tablenum{	69}	&\tablenum{	67	}&	\tablenum{75}	&	\tablenum{63	}\\ 
\hline
&$\mathrm{^{13}C}$-6&$\mathrm{^{13}C}$-7&$\mathrm{^{13}C}$-8&$\mathrm{^{18}O}$&\\
\hline
$A$ /MHz
 &     \tablenum{   4345.7658(16)}

 &  \tablenum{ 4334.2406(14)}

 &  \tablenum{   4326.5526(12)}

&\tablenum{4312.4667(25) }
&
 \\ 
%\hline 
$B$ /MHz
&
    \tablenum{  1124.76691(50)}
&
 \tablenum{ 1118.62448(45)}
&
  \tablenum{    1107.35228(35)}
&
\tablenum{   1093.80668(77)  }
&
\\ 
%\hline 
$C$ /MHz
&
   \tablenum{    951.00022(44) }
 & 
 \tablenum{   946.13750(36) }
&
 \tablenum{      939.70914(28) }
&
\tablenum{   929.40768(52) }
 &
 \\ 
$\sigma$ /kHz  &  \tablenum{44.6 }	&	\tablenum{20.3}	&	\tablenum{ 24.2	}&	\tablenum{21.8} & 	\\ 
No. of lines & 	\tablenum{	37	}&	\tablenum{61	}&	\tablenum{69} &\tablenum{ 35}&\\ 
\hline
%\hline
%%%%%%%%%%%%%%%%%%
\end{tabular} 
\end{small}
\label{isotopologues}
\end{table*}

The corresponding rotational constants are summarised in Table \ref{isotopologues}. 
The centrifugal distortion constants of the parent species were used for the assignment of the rare $\mathrm{^{13}C}$ and $\mathrm{^{18}O}$ isotopologues in natural abundance.
Independent fits of the quartic centrifugal distortion constants yielded only a 2 kHz frequency uncertainty improvement (in average) at the cost of much higher errors for the constants. 
This can be explained by the low number of lines (between 35 and 78) and the low frequency range with lower $J$ and $K$ values that were assigned for these species.
\\
Using the rotational constants for the singly-substituted isotopologues, we employed the substitution ($\mathrm{r_s}$) method to extract the carbon and oxygen atomic coordinates with respect to the principal axes of the molecule using Kraitchman's equations \citep{Kraitchman.1953}.
In Figure \ref{structure}, we compare the structural parameters of (R)-(+)-styrene oxide obtained using the $\mathrm{r_s}$ method with the $\mathrm{r_0}$ method \citep{Kisiel.2003}, where the effective structure of the ground state is partially determined by a least squares fit including all the rotational parameters. Relevant bond lengths and angles are shown in the figure and compared with the calculated values. The results of the B3LYP/def2-TZVP calculation are overlayed with the $\mathrm{r_s}$ structure, represented by the solid spheres. 
\begin{figure*}[ht]
\centering
\includegraphics[scale=0.65]{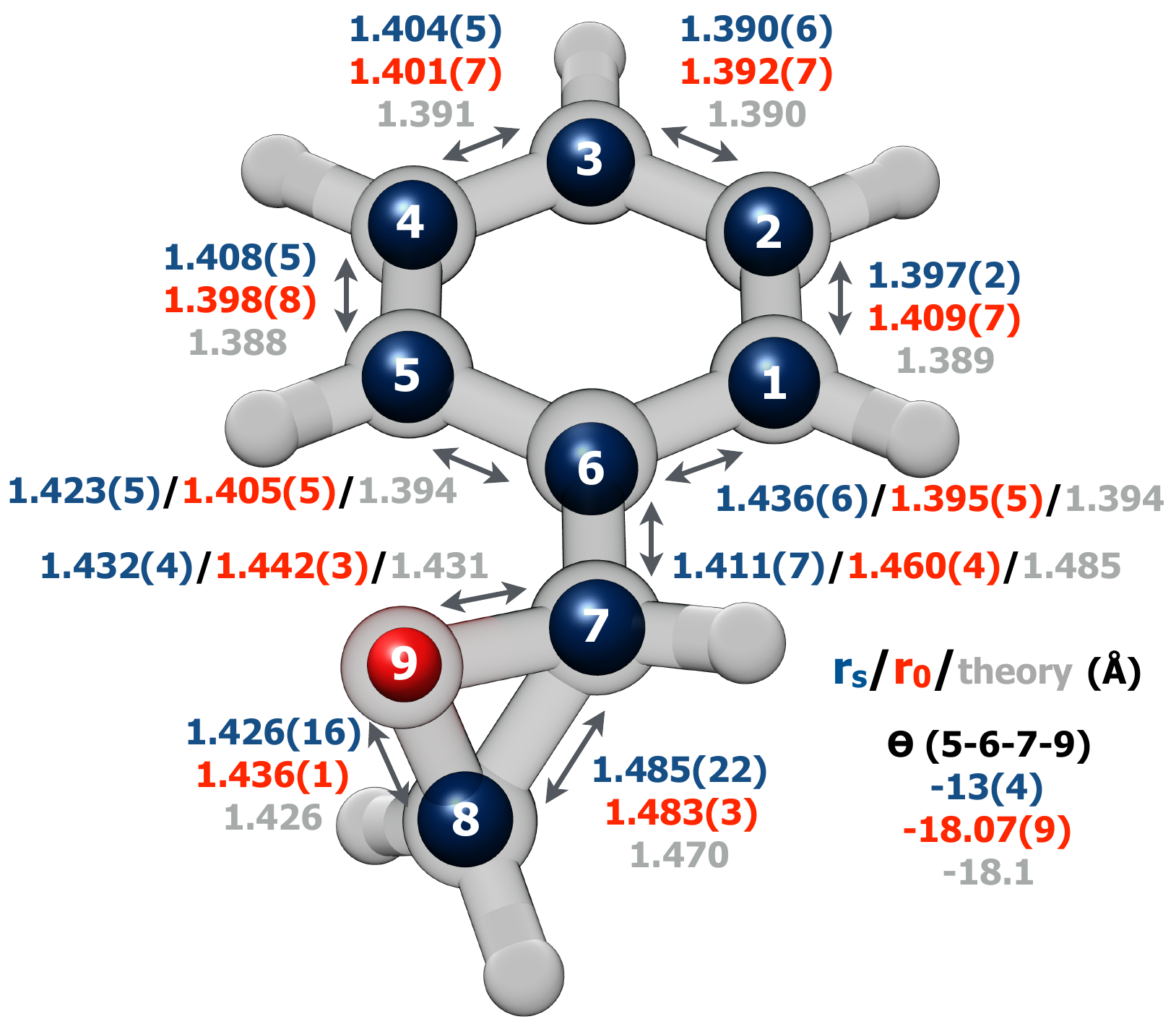}
\caption{Overlay of the experimental and predicted (B3LYP/def2-TZVP) structure of SO. The solid spheres represent the experimentally ($\mathrm{r_s}$) determined carbon and oxygen atom coordinates. Relevant bond lengths in $\si{\angstrom}$ and angles in degrees are shown for (R)-(+)-styrene oxide.
Experimental bond lengths from the $\mathrm{r_s}$ and $\mathrm{r_0}$ structures are depicted  with their uncertainties.
}
\label{structure}
\end{figure*}

\subsection{Sources for astronomical searches} \label{astro} 
The combined laboratory and theoretical efforts allow a precise prediction of transition frequencies of styrene oxide in its ground state at wavelengths that overlap with the wavelength bands of the Effelsberg telescope, GBT, and ALMA observatory (see Figure \ref{fig:bands}). SO can possibly be found in cold, dark clouds, as well as in hot core regions close to newly born stars. In the latter case, temperatures can reach up to 100 K or above, and the spectrum of styrene oxide is predicted to be strongest in the ALMA bands 5 and 6, see Figure \ref{fig:bands}. For lower temperatures, ALMA band 3 is more suitable, which widely overlaps with the W-band spectrometer's frequency coverage, and direct laboratory data is available. Furthermore, the Effelsberg telescope band around 35 GHz is most suitable with regard to lower temperatures. This is also true for the GBT, which covers both the low frequency range and frequencies up to 110 GHz, including the W-band. The GBT is able to address promising sources, e.g. TMC-1 and Sgr B2(N), in particular as benzonitrile and methyloxirane have been found in these sources \citep{McGuire.2016b, McGuire.2018b}.
\\
As outlined above, our measurements cover a wide wavelength range up to sub-millimeter wavelengths (sub-mmw). This enabled us to determine the molecular distortion constants with high precision and to make frequency predictions into the sub-mmw range, i.e., up to several hundred GHz.   
Our line predictions of the ground state (a line list can be found in the ESI\dag~ as a .cat file) are based on the parameters given in Table \ref{tab:results}, and the corresponding partition functions $Q$ for the Hamiltonian with $J_{max}$=120 for different temperatures are given in Table S4 of the ESI\dag. 
\\
Due to the possibility of line confusion caused by the dense spectra of COMs in the mmw and sub-mmw region, also known as the interstellar weeds that are related to hot cores \citep{Herbst.2009b}, it is also important to not only know the ground state but also the vibrationally excited states of a molecule.
Thus, in the possible presence of SO in astronomical sources, our prediction up to sub-mm wavelengths for $v_{45}$, $v_{44}$, and $v_{43}$ might help to assign and link their strong rotational transitions to this molecule in the ISM.  
\\
The additional detection of $\mathrm{^{13}C}$-containing isotopologues of styrene oxide could help to further elucidate the chemical formation route of this molecule. As has been illustrated, by applying a numerical gas-grain chemical network model of TMC-1, differences in the isotopic fractionation of CCH and CCS occur depending on whether these molecules have been formed by CO or a single C atom \cite{KenjiFuruya.2011}. In a similar manner, the knowledge of the isotopic ratio and the position of $\mathrm{^{13}C}$ within styrene oxide could be used to deduce possible precursors and thus help to pin down the formation route of this molecule.

\section{Conclusion and Outlook}
In this work, we present a broadband high-resolution spectral analysis of the chiral molecule styrene oxide between 2 GHz and 330 GHz. We provide accurate predictions of transition frequencies of this molecule up to (sub-)millimeter wavelengths. The fitted rotational transitions of the vibrational ground state exhibit an average uncertainty of about 54 kHz including transitions up to $J$=120. We determined the rotational constants and centrifugal distortion constants up to the sextic order. We also determined the rotational parameters for the three normal modes $v_{45}$, $v_{44}$, and $v_{43}$, and we were able to assign rotational transitions originating from these three vibrationally excited states up to sub-mm wavelengths. Furthermore, we assigned all singly-substituted $\mathrm{^{13}C}$ and $\mathrm{^{18}O}$ isotopologues of styrene oxide and determined the rotational constants for each species. Using this data, we determined the gas phase structure of styrene oxide. Our data will assist the astronomical search for styrene oxide in the ISM. 
Potential molecular sources of SO are either hot cores or cold clouds. However, there are a few constraints to be considered. At very low temperatures (few K), styrene oxide might be locked on icy dust grains where it might have been formed. Thus, a gas-phase detection at these temperatures might be unlikely. For environments with temperatures exceeding 150 K to 200 K, the partition function becomes large, and the intensity of individual lines will decrease strongly. Consequently, sources with temperatures in between these two extremes could be appropriate. Examples of sources that might be suitable for a styrene oxide detection are the star-forming region TMC-1 and Sgr B2(N) towards the galactic center. These sources are promising because in TMC-1, the aromatic molecule benzonitrile has been found \citep{McGuire.2018b}, and in Sgr B2(N), methyloxirane, the first chiral molecule detected in space, was observed \citep{McGuire.2016b}. Since SO consists of a phenyl ring attached to an oxirane group, these sources' characteristics can lead to either a search for SO in one of these, or other objects with similar chemical conditions. The Effelsberg telescope, GBT, and ALMA work at wavelengths that perfectly overlap with our measured wavelength range. An astronomical detection of styrene oxide would help to better understand the formation pathways of chiral molecules and possibly those of their precursors.
With the molecular constants given in Table \ref{isotopologues}, transition frequencies of styrene oxide isotopologues can be calculated, enabling an astronomical search for these important heavy-atom isotopologues and possible formation routes of SO.

\section{Acknowledgements}
TFG, MS, and PS acknowledge the funding by the Deutsche Forschungsgemeinschaft (DFG, German Research Foundation) – Projektnummer 328961117 – SFB 1319 ELCH.  PS and GWF also gratefully acknowledge the funding from the DFG-FU 715/3-1 project. MS acknowledges the funding by the ERC Starting Grant ASTROROT (638027).
BEA acknowledges funding from the International Max Planck Research School for Ultrafast Imaging and Structural Dynamics (IMPRS-UFAST).

\section*{Conflicts of interest}
%In accordance with our policy on \href{http://www.rsc.org/journals-books-databases/journal-authors-reviewers/author-responsibilities/#code-of-conduct}{Conflicts of interest} please ensure that a conflicts of interest statement is included in your manuscript here.  Please note that this statement is required for all submitted manuscripts.  If no conflicts exist, please state that 
There are no conflicts to declare.
%%%END OF MAIN TEXT%%%

%  For footnotes in the main text of the article please number the footnotes to avoid duplicate symbols. e.g.  \footnote[num]{your text} the corresponding author \ast counts as footnote 1, ESI as footnote 2, e.g. if there is no ESI, please start at [num]=[2], if ESI is cited in the title please start at [num]=[3] etc. Please also cite the ESI within the main body of the text using \dag.

% The \balance command can be used to balance the columns on the final page if desired. It should be placed anywhere within the first column of the last page.

% \balance

% If notes are included in your references you can change the title from 'References' to 'Notes and references' using the following command:
\renewcommand\refname{References}

%%%REFERENCES%%%
\scriptsize{
%\bibliography{C8H8Onew_10}
\providecommand*{\mcitethebibliography}{\thebibliography}
\csname @ifundefined\endcsname{endmcitethebibliography}
{\let\endmcitethebibliography\endthebibliography}{}

\bibliographystyle{rsc}  } %the RSC's .bst file

\end{document}